\pdfoutput=1 
\documentclass[amsmath,amssymb,floatfix,prl,twocolumn]{revtex4}
\usepackage[pdftex]{graphicx}
\usepackage{amssymb,amsfonts,amsmath}
\usepackage{graphicx,amsmath}
\usepackage{float}
\usepackage{color}
\usepackage{siunitx}
\def\s{\sigma}

\begin{document}
\bibliographystyle{unsrt}
\title{Efficient charge separation in organic photovoltaics through incoherent hopping}
\author{Stavros Athanasopoulos}
\email{astavros@fis.uc3m.es}
\thanks{}
\affiliation{Departamento de F\'isica, Universidad Carlos III de Madrid, Avenida Universidad 30, Legan{\'e}s 28911, Madrid, Spain}
\author{Steffen Tscheuschner}
\affiliation{Experimental Physics II, University of Bayreuth, Bayreuth 95440, Germany}
\author{Heinz B{\"a}ssler}
\affiliation{Bayreuth Institute of Macromolecular Research (BIMF), University of Bayreuth, Bayreuth 95440, Germany}
\author{Anna K{\"o}hler}
\affiliation{Experimental Physics II, University of Bayreuth, Bayreuth 95440, Germany\\
Bayreuth Institute of Macromolecular Research (BIMF), University of Bayreuth, Bayreuth 95440, Germany}

\begin{abstract}
We demonstrate that efficient and nearly field-independent charge separation in organic planar heterojunction solar cells can be described by an incoherent hopping mechanism. We model the separation efficiency of electron-hole pairs created at donor-acceptor organic interfaces. By using kinetic Monte Carlo simulations that include the effect of on-chain delocalization we show that efficient charge extraction to the electrodes requires bipolar transport and increased dimensionality. This model explains experimental results of almost field independent charge separation in some molecular systems and provides important guidelines at the molecular level for maximizing the efficiencies of organic solar cells.
\end{abstract}

\maketitle

Charge separation in solar cells is a key process for extracting carriers that contribute to a photocurrent. While in traditional inorganic solar cells the available thermal energy provides enough kinetic energy in order to overcome the Coulomb interaction between geminate electron-hole pairs, in new generation organic solar cells, separation of charges is a cumbersome process~\cite{Mihailetchi2004, AnnaHeinzBook}. As a first step, it usually requires the presence of interfaces between an electron transporting and a hole transporting material that facilitates the transfer of the electron to the acceptor material by formation of a charge transfer state~\cite{Sariciftci1992}, provided that the exciton diffusion length is large enough to reach such interfaces~\cite{PRB2009,JPCC2008,Mikhnenko2015}. Second, once the charges are in separate phases, they need to overcome their mutual Coulomb attraction that is much larger than the thermal energy, due to the low dielectric constant of organic materials, which is in the range of 3-5~\cite{Deibel2010}. This charge separation process can be assisted by an electric field, although in some materials the process occurs with high efficiency even at low electric fields. A great deal of attention has been drawn recently on the mechanism of charge separation and a number of possible scenarios have been suggested to explain the large variability in dissociation efficiencies for chemically similar molecular systems~\cite{Few2015}. These include dissociation via delocalized or hot states~\cite{Nayak2013,Peumans2004,Clarke2010}, ultrafast separation via band-type states~\cite{Gelinas2014,Savoie2014}, the presence of interfacial dipoles~\cite{Schwarz2013,Tscheuschner2015,Verlaak2009}, entropic contributions~\cite{Gregg2011,Gao2015} and coherence effect~\cite{Bittner2014}.

The aim of this Letter is to report on the role of diffusion on the sensitive balance between the extraction of photogenerated charges and their recombination. In organic solar cells both, geminate recombination, i.e. the recombination of electron-hole carriers that trace back to the same parent exciton, and non-geminate (bimolecular) recombination, i.e. recombination of electron-hole carriers from two different parent excitons, are important loss mechanisms that reduce the quantum efficiency of the device. It has recently been demonstrated that geminate recombination in small molecule based solar cells can be controlled by thermal annealing and chemical treatment of spin coated devices with the magnitude being accessible experimentally by time-delayed collection field measurements (TDCF)~\cite{Proctor2014}. Here we treat the problem of geminate recombination versus charge extraction from a simulation standpoint. In this Letter we shed light on the mechanism of charge extraction in bilayer donor-acceptor devices by using Monte Carlo simulations  ~\cite{Peumans2004,Watkins2005,Marsh2007,Groves2008} to calculate the charge collection efficiency as a function of externally applied field. We lay emphasis on the role of dimensionality, disorder and bipolar transport. In addition, we examine the question of delocalization, whereby the spatial extent of a charge wavefunction over a segment is parameterized by an effective mass of the polymer. 

\begin{figure}
\centerline{\includegraphics[clip=true,width=8cm]{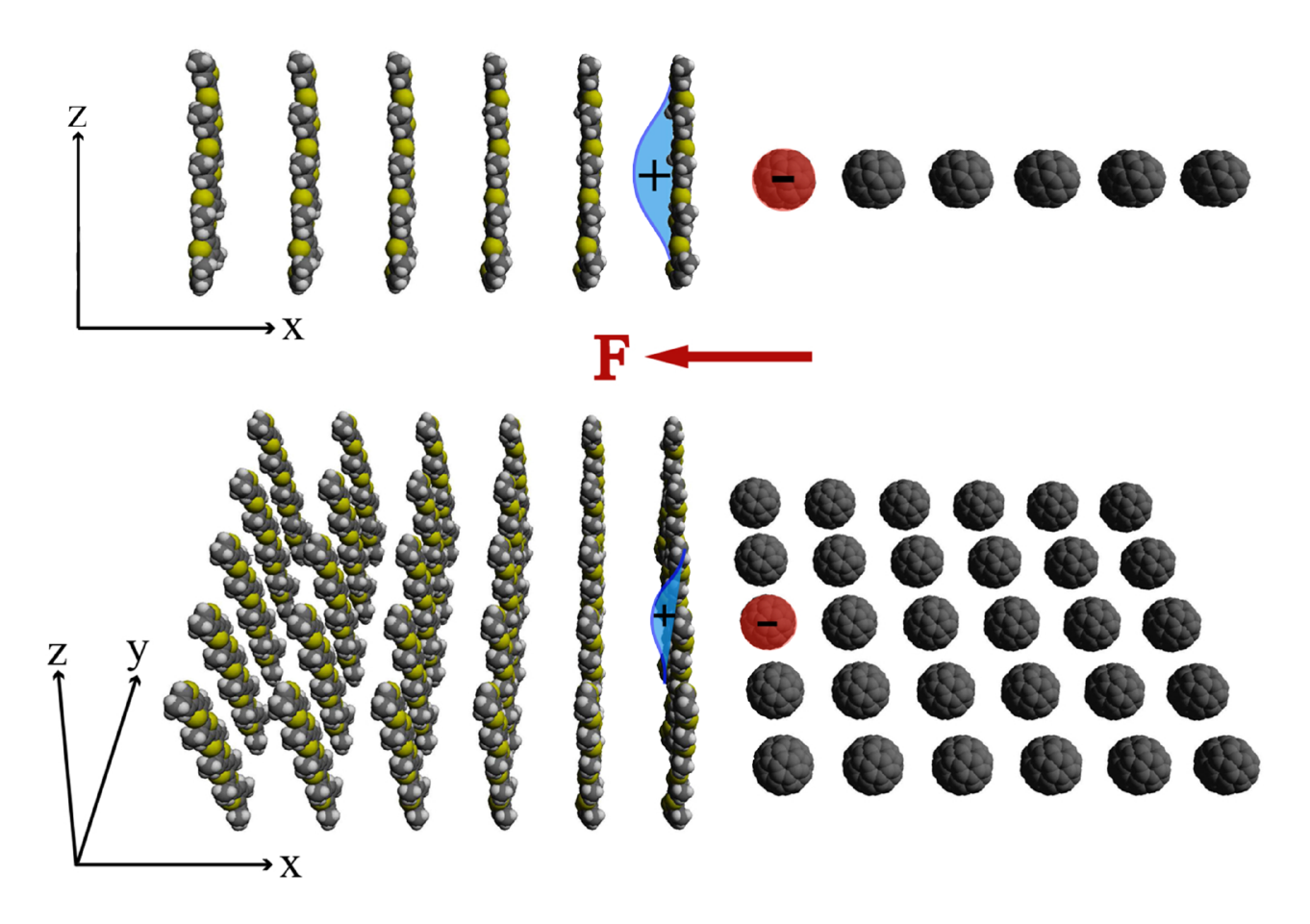}}
\caption{(Color online) Schematics of the 1D (top panel) and 2D (bottom panel) bilayer lattice morphologies used for the Monte Carlo simulations. Each grid point consists of either a donor polymer chain of infinite length or a C$_{\text{60}}$ acceptor molecule. The effect of on-chain hole delocalization is taken into account using an effective mass model. 
}
\label{Fig1}
\end{figure}

In what follows, we compute the probability that an electron-hole pair separates over a given distance of  \SI{100}{\nm}. We use a discrete lattice of points with donor polymers as the basis with their molecular axis extending parallel to each other and to the interface with the acceptor. The hole can only make inter-chain hops between $\pi$-conjugated polymers. The extent of the hole wavefunction along the polymer chain is taking into account using an effective mass model~\cite{Tscheuschner2015,Arkhipov2003,Nenashev2011}. The main idea of the model is that the zero-point oscillations of the delocalized hole modify the potential energy of the hole by an additional kinetic energy term in the Hamiltonian. The potential energy of the hole is determined by the Coulomb potential well due to the presence of the electron and the external electric field. The more delocalized the hole is, the larger the kinetic energy term and therefore the shallower the potential. The degree of delocalization can be described by an on-chain effective mass for the hole $m_{eff}$, given in units of the free electron rest mass.

We consider a linear 1-dimensional (Figure 1 top panel) and a square 2-dimensional lattice (Figure 1 bottom panel) with a constant $a= \SI{1}{\nm}$. For simplicity the lattice constant remains the same for the acceptor phase, representing a buckyball based molecule, and energetic disorder is neglected. The initial condition for each independent Monte Carlo trial sets a photogenerated electron-hole pair at the interface, in a nearest neighbor charge transfer configuration with an initial separation of \SI{1}{\nm}. Since we focus on geminate recombination, at every Monte Carlo trial we follow only one pair of charges. An external electric field of strength F is applied perpendicular to the DA interface with a vector direction antiparallel to the electron-hole Coulomb field. 

Charge hopping rates follow the Miller-Abrahams formulation~\cite{Miller1960}:

\begin{equation} 
\nu_{ij}=\left\{ \begin{array}{ll}
 \nu_{0}e^{-2\gamma r_{ij}}e^{\left(-\frac{\varepsilon_j-\varepsilon_i}{k_BT}\right)} &\mbox{ for $\varepsilon_j>\varepsilon_i$} \\
 \\
  \nu_{0}e^{-2\gamma r_{ij}} &\mbox{ for $\varepsilon_j\leq \varepsilon_i$}
       \end{array} \right.
\label{eq:MArates}
\end{equation}

where $i$ denotes the residence site of the charge and $j$ the target site, and the two sites are separated by a distance $r_{ij}$. Site energies $\varepsilon_i$ and $\varepsilon_j$ include contributions from the electron-hole Coulomb interaction potential including the kinetic energy term resulting from the delocalization of the hole wavefunction calculated by numerically solving the Schr{\"o}dinger equation (see Ref.~\cite{Tscheuschner2015}),  the voltage drop due to the applied field and the static Gaussian disorder~\cite{Bassler1993}. Unless stated, the inverse localization length $\gamma=\SI{2}{\nm^{-1}}$, the frequency prefactor \mbox{$\nu_0=10^{12}$ {\SI{}{\s^{-1}}}} and $T= \SI{300}{\K}$ . We consider hopping events up to second nearest neighbor distances, ie $r_{ij,max}=\SI{2}{\nm}$, which is a sufficient cut-off distance for moderate values of temperature normalized disorder $\sigma/kT<4$. From a simulation point of view, the parameters $\nu_0$ and $\gamma$ define the minimum hopping time and their value does not influence the presented results since it is the relative ratio of the CT lifetime to the minimum hopping time that controls the probability for recombination (see below). In practice, these parameters determine the equilibrium charge carrier diffusion coefficient and mobility values.

At each Monte Carlo step we calculate a waiting time for each hopping event: $\tau_{ij}=-(1/\nu_{ij})lnX$. In addition, we calculate a waiting time for recombination events between the electron-hole pair: $\tau_r=-\tau lnX$, where $\tau$ is the electron-hole pair lifetime that increases exponentially with electron-hole distance $r_{eh}$ as $\tau=\tau_{CT} e^{2\gamma(r_{eh}-a)}$  and $X$ is a random number drawn from a box distribution between 0 and 1. The lifetime at close proximity is a parameter that we allow to vary from a minimum value of $\tau_{CT}=35t_0$ to a maximum value of $\tau_{CT}=3000t_0$, with $t_0$ being the minimum hopping time $t_0=\frac{1}{\nu_0}  e^{2\gamma a}$. The event with the smallest waiting time is selected and executed and the interaction potential updated. If the accepted event is a hop, then we update the site of the hole or electron and recalculate waiting times. If the chosen event is recombination, we remove the charges and start a new trial. Each trial terminates successfully when the electron-hole distance is larger or equal to \SI{100}{\nm}. The influence of the donor acceptor thickness has been considered elsewhere~\cite{Tobias2016}. The statistical quantity of interest is the separation yield calculated as: $\varphi(F)=\frac{N_{sep}(F)}{N_{tot}(F)}$, where $N_{sep}(F)$ the number of successful trials (trials with $r_{eh}\geq \SI{100}{\nm}$) for an applied field $F$ and $N_{tot}(F)$ the total number of trials of the order of $10^4$. 

\begin{figure}
\centerline{\includegraphics[clip=true,width=8cm]{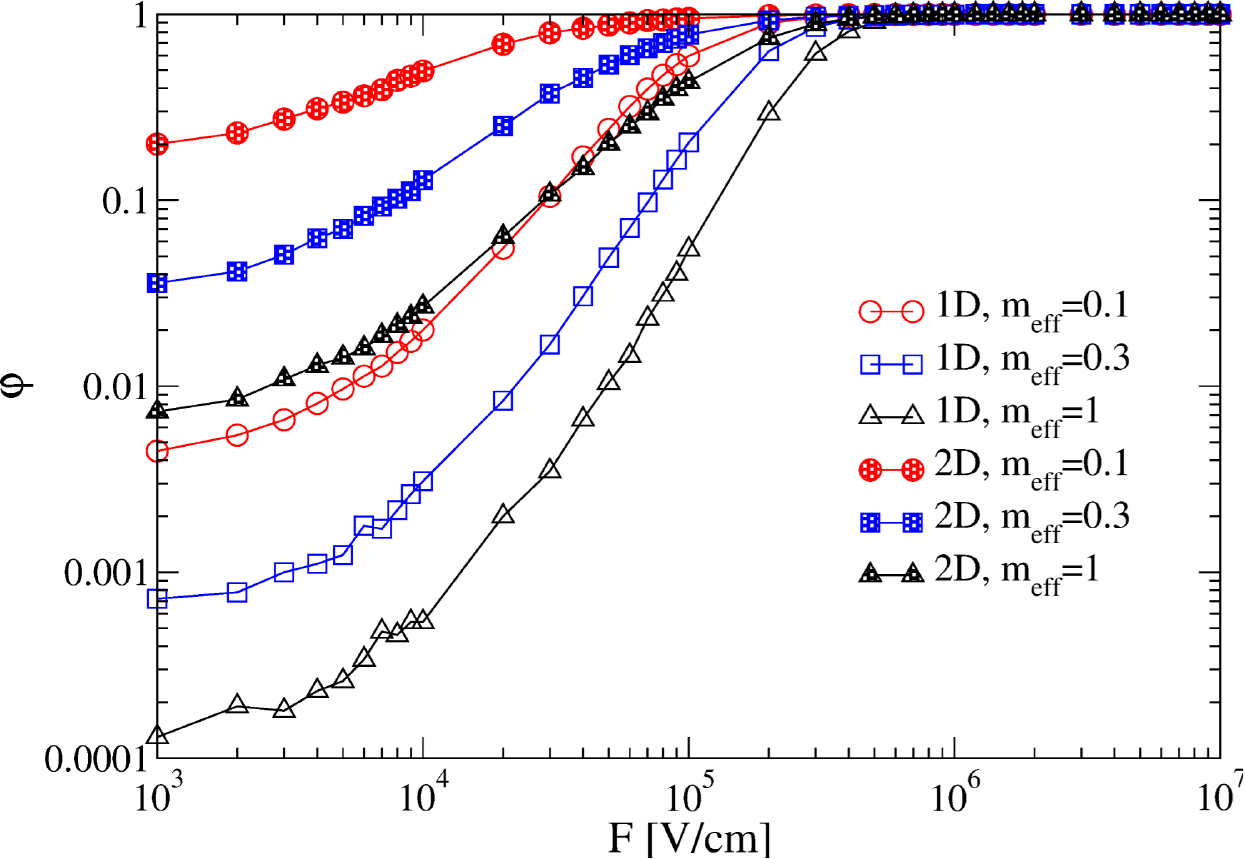}}
\caption{(Color online) 1D and 2D Monte Carlo simulations for different effective mass of the donor molecules. Electron extraction efficiency φ as a function of an internal electric field F. 
}
\label{Fig2}
\end{figure}

At first we consider that only one of the carriers is mobile, e.g. only the electron is allowed to hop and the hole position is fixed. This situation could also be modeled analytically in one dimension~\cite{Rubel2008} and we have tested that the results for the field dependence of the separation yield are in perfect agreement with the analytical solution. Figure 2 shows that reducing the effective mass of the donor results in higher electron extraction efficiencies at intermediate and low fields. This is a trivial result, because a lower effective mass, i.e. a longer hole delocalization along the polymer chain, translates into a smaller barrier for the e-h interaction potential (see Ref~\cite{Tscheuschner2015}).  Let us now consider the 2-dimensional case where the electron has more routes available to escape the e-h interaction potential while the hole is still stationary at the interface. Simulations reveal that $\varphi$ increases by more than one order of magnitude for $m_{eff}=0.1m_e$ at low fields where the driving force for electron extraction is dominated by the diffusion of the electron rather than the electric field (Fig. 2).  This is a remarkable result that highlights the important role of dimensionality on diffusion limited charge extraction.

 \begin{figure}
\centerline{\includegraphics[clip=true,width=9cm]{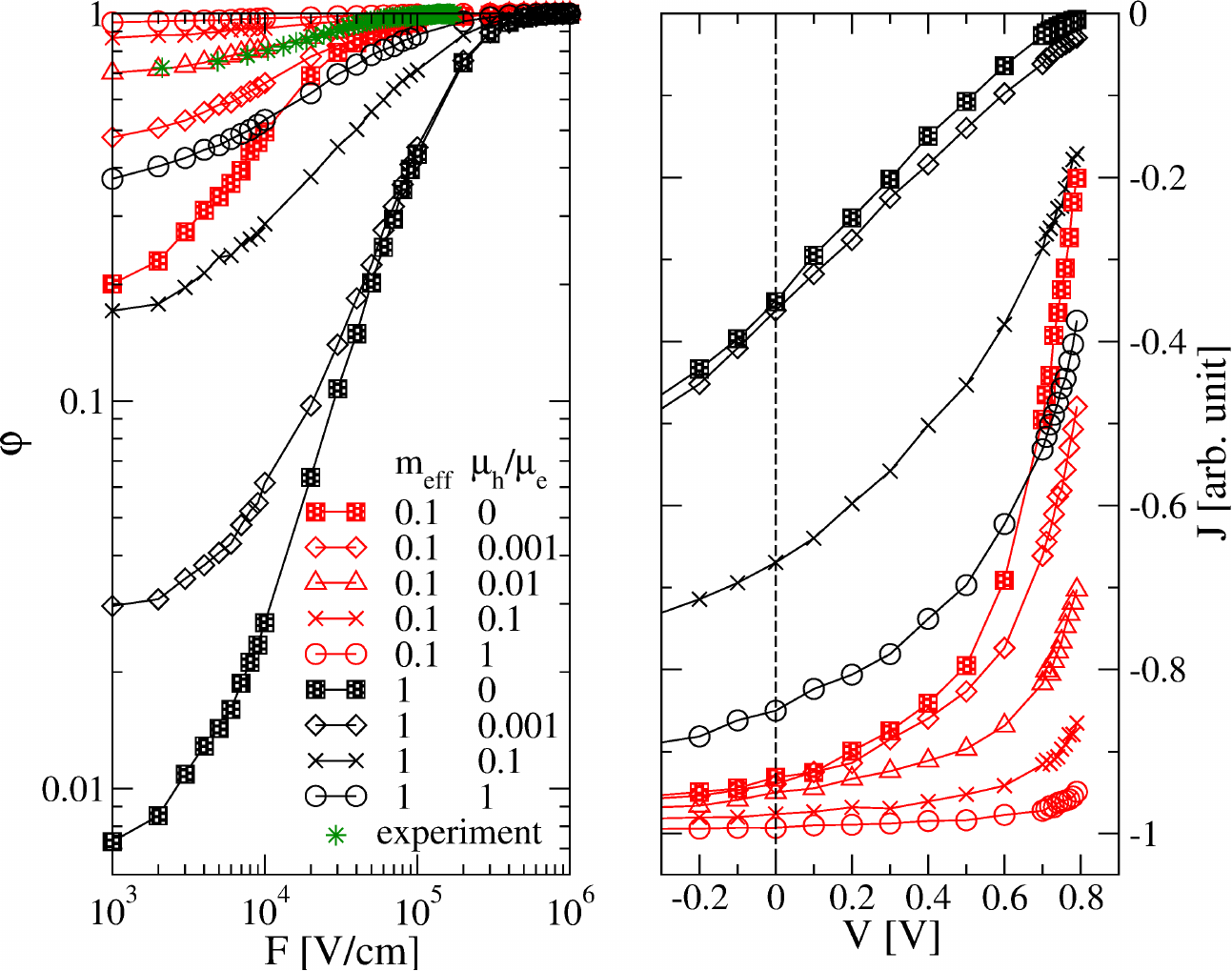}}
\caption{(Color online) 2D Monte Carlo simulations, influence of imbalanced e and h mobilities. Left panel: Electron-hole separation efficiency φ as a function of an externally applied field and experimental data for a bilayer consisting of \SI{60}{\nm} thick layer of the T1 molecule as donor and \SI{30}{\nm} thick layer of C$_{\text{60}}$ as acceptor. Right panel: resulting photocurrent JV curves for a purely geminate recombination regime.
}
\label{Fig3}
\end{figure}

Let us now lift up the restriction of immobilizing the hole and further allow for imbalanced electron and hole mobilities. Results from the Monte Carlo calculations in the 2D case depicted in Figure 3 show a striking increase in the e-h separation yield when both the hole and the electron particle are allowed to hop. For a hole effective mass $m_{eff}=0.1$, $\varphi$ becomes virtually field independent even at the diffusive regime at low electric fields. We note that there are several reports of organic solar cells with molecular oligomers as the donor material that exhibit high external quantum efficiencies~\cite{Proctor2014,Zhang2015,Kan2014} and we have recently reported high separation yields in bilayer cells using the p-DTS(FBTTh$_{\text{2}}$)$_{\text{2}}$ oligomer (T1 molecule) as donor with C$_{\text{60}}$ as acceptor~\cite{Tobias2016}. These data are included at the left panel of Figure 3 and it is noteworthy that the field dependence can be well reproduced from the Monte Carlo simulations for $m_{eff}=0.1$ and a mobility imbalance ratio $\mu_h/\mu_e=10^{-2}$. Such a ratio is consistent with photo-CELIV measurements of hole mobility $1.4\pm0.5*10^{-4}$ \si{\cm^2/Vs} for the T1 oligomer while the electron mobility in C$_{\text{60}}$ is of the order of $10^{-2}$ \si{\cm^2/Vs}.~\cite{Tobias2016} We can further translate the obtained Monte Carlo results to photocurrent JV curves, assuming a thickness of the active layer (electron and hole transporting material) of $d=\SI{100}{\nm}$  in total, and an open-circuit voltage $V_{oc}=\SI{0.8}{\V}$. The simulated JV characteristics are depicted at the right panel of Figure 3. They represent purely geminate recombination limited JV curves. These results highlight that bipolar transport dramatically increases the fill factors (FF) by lowering recombination, even up to the limit where for a system with low effective mass and balanced electron and hole transport ideal JV characteristics are recovered. The geminate recombination regime can be partly accessed experimentally via time-delayed collection field measurements~\cite{Proctor2014}.  Experiments in as-cast and thermally annealed T1:PC$_{\text{70}}$BM film blends have shown that annealed films exhibit phases with higher crystalline domains of T1 that improve hole mobilities by an order of magnitude as measured by photo-CELIV and TDCF transients. Consequently, the fill factors for annealed films have been shown to improve. This is therefore in accordance with the predictions of the Monte Carlo model.   

\begin{figure}
\centerline{\includegraphics[clip=true,width=8cm]{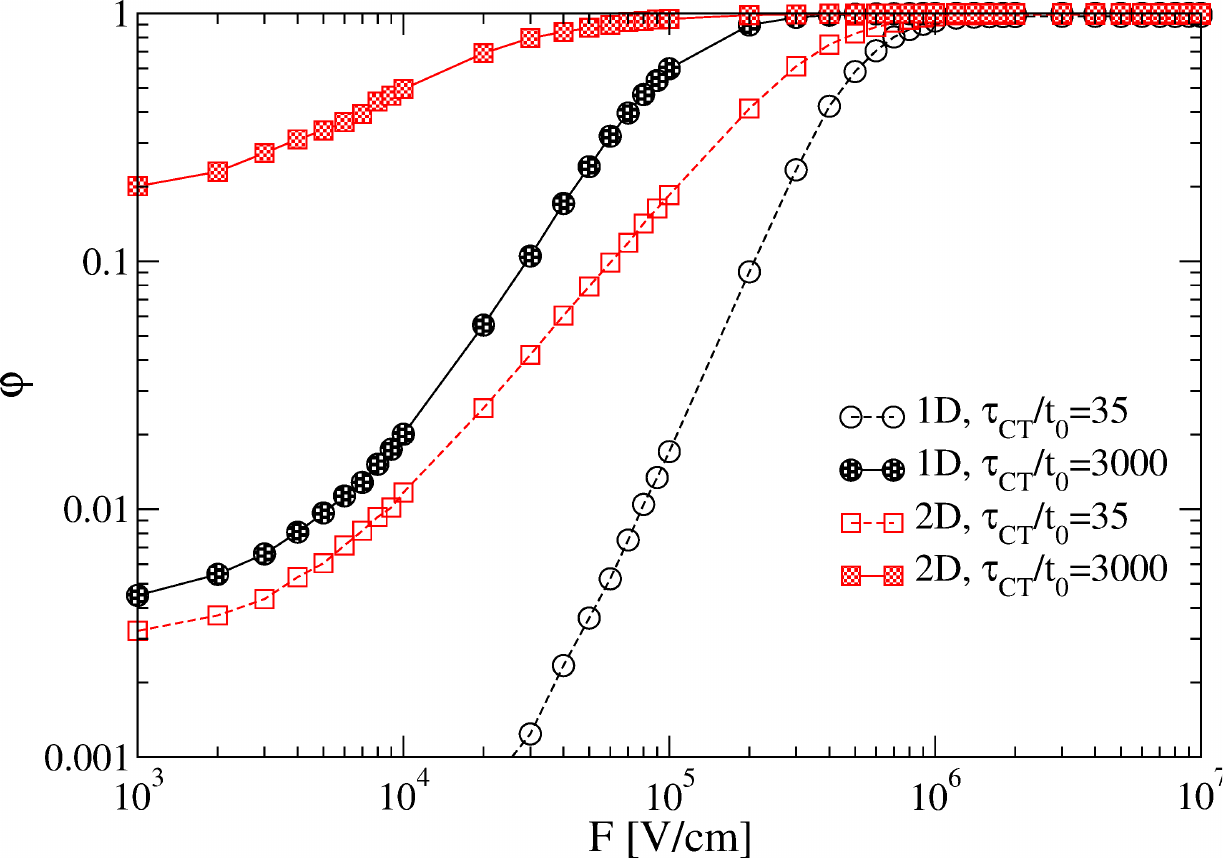}}
\caption{(Color online) 1D and 2D Monte Carlo simulations for $m_{eff}=0.1$ and different CT state lifetime over minimum hopping time ratios: Electron-hole separation efficiency $\varphi$ as a function of an  internal electric field F.
}
\label{Fig4}
\end{figure}

Next, we consider the influence of varying the charge transfer state lifetime with respect to the minimum hopping time that determines the magnitude of the mobility of the particles. This can be achieved by either increasing the charge transfer state lifetime of the pair, which corresponds to decreasing the electron-hole wavefunction overlap, or by decreasing the minimum hopping time, equivalent to increasing the coupling between the molecular units. When the ratio $\tau_{CT}/t_0$ of the CT state lifetime over minimum hopping time increases, this corresponds to increasing the $\mu\tau$ product in a Braun-Onsager model~\cite{Braun1984}. Consequently, the separation yield increases since the electron and hole are given more chances to make hops and can get far-off in the course of their lifetime. Figure 4 shows the changes in $\varphi$ for the 1D and 2D cases. At the low field regime, $\varphi$ increases by the same order of magnitude that  $\tau_{CT}/t_0$ increases. The role of energetic disorder has also been investigated and we find that the inclusion of Gaussian disorder does not alter the conclusions presented here. A detailed study on the influence of disorder will be presented elsewhere.

We summarize the main results obtained in this Letter. We are not only able to simulate experimental photocurrent JV curves dominated by geminate recombination, but we can also identify several parameters, which all independently lead to higher charge separation yields with weaker field dependence. These parameters are: (i) increase in transport dimensionality, (ii) bipolar transport, (iii) low effective mass and, conventionally known, (iv) increase in the minimum hopping time over CT state lifetime ratio, i.e. increasing the $\mu\tau$ product. Given that typically CT state lifetimes do not vary dramatically in most organic blends overall (i) has the strongest effect following by (ii) and (iii).   For (ii) we find that fully balanced mobilities is less important~\cite{Tress2011}, which is promising given that typically electron mobilities can be one or two orders of magnitude higher than hole mobilities. (iii) is mainly controlled by molecular design with $m_{eff}$ varying from about 0.1 in well conjugated polymers to values of the order of 0.3 in oligomers and values of 1 in disordered molecules exhibiting strong hole localization. 
We stress that our results are also relevant to recent experiments that show that either delocalized hole states or fullerene aggregation improves charge separation~\cite{Gelinas2014,Bassler2015}.  In fact, in comparison with published experimental results, our simulations explain the increase of fill factors with increased crystallinity obtained by thermal annealing or chemical treatment. The delocalization of the hole wavefunction assists the transport of the electron in the acceptor phase by lowering the Coulomb interaction potential, whereas fullerene aggregation is expected to increase hopping rates. Our findings have also the following consequences for optimizing organic solar cell devices: Firstly, for the design of organic solar cells, it is not sufficient to focus on only one material displaying delocalized states. Rather, both materials need to display either high charge delocalization or high carrier mobilities, ideally both. Secondly, morphologies that promote filamentary transport are detrimental for charge separation, whereas architectures that favor transport with increased dimensionality (even at the expense of localization) are superior.  It is also important to highlight that our results were obtained within an incoherent hopping formalism of vibrationally relaxed CT states~\cite{Bassler2015} where the initial electron-hole pair has a close proximity of only \SI{1}{\nm}. Therewithal, we find that an incoherent transport model can describe efficient charge separation without the need to invoke "hot" processes, in agreement with~\cite{Vandewal2014}.

\acknowledgments This project has received funding from the Universidad Carlos III de Madrid, the European Union’s Seventh Framework Programme for research, technological development and demonstration under grant agreement n$^{\text{o}}$ 600371, el Ministerio de Econom{\'i}a y Competitividad (COFUND2014-51509), el Ministerio de Educaci{\'o}n, cultura y Deporte (CEI-15-17) and Banco Santander. We also acknowledge additional funding from the German Research Foundation DFG (GRK1640) and the Bavarian Academic Centre for Latina America (BAYLAT).

\bibliography{arxiv_CS}

\end{document}